\documentclass[runningheads]{llncs}

\usepackage[utf8]{inputenc} 
\usepackage[T1]{fontenc}    
\usepackage{hyperref}       
\usepackage{url}            
\usepackage{booktabs}       
\usepackage{amsfonts}       
\usepackage{nicefrac}       
\usepackage{microtype}      
\usepackage{xcolor}         

\usepackage{graphicx}
\usepackage{comment}
\usepackage{amsmath,amssymb} 
\usepackage{color}
\usepackage{wrapfig}
\usepackage{enumitem}
\usepackage{cleveref}

\usepackage{multirow}
\usepackage{subfigure}
\usepackage{algorithmic}
\usepackage[ruled,vlined]{algorithm2e}
\usepackage{lipsum}
\usepackage{stackengine}
\newcommand{\myparagraph}[1]{\smallskip\noindent\textbf{#1}}
\newcommand{\method}{\textit{P-Count}}

\DeclareMathOperator*{\argminA}{arg\,min} 

\DeclareMathOperator*{\argmaxA}{arg\,max} 

\begin{document}
\title{\method{}: Persistence-based Counting of White Matter Hyperintensities in Brain MRI}
%
%
%

\author{Xiaoling Hu\inst{1\star}\and
Annabel Sorby-Adams\inst{1} \and
Frederik Barkhof\inst{2,3} \and
W Taylor Kimberly\inst{1} \and
Oula Puonti\inst{1,4} \and
Juan Eugenio Iglesias\inst{1,2,5}}

\renewcommand{\thefootnote}{\fnsymbol{footnote}}

\footnotetext[1]{Corresponding to: Xiaoling Hu (xihu3@mgh.harvard.edu).}
\institute{Martinos Center for Biomedical Imaging, MGH and Harvard Medical School \and Center for Medical Image Computing, University College London \and Amsterdam University Medical Center \and Danish Research Centre for Magnetic Resonance, Copenhagen University Hospital \and Computer Science and Artificial Intelligence Laboratory, MIT}

\maketitle              
\begin{abstract}
White matter hyperintensities (WMH) are a hallmark of cerebrovascular disease and multiple sclerosis. Automated WMH segmentation methods enable quantitative analysis via estimation of total lesion load, spatial distribution of lesions, and number of lesions (i.e., number of connected components after thresholding), all of which are correlated with patient outcomes. While the two former measures can generally be estimated robustly, the number of lesions is highly sensitive to noise and segmentation mistakes -- even when small connected components are eroded or disregarded. In this article, we present \method{}, an algebraic WMH counting tool based on persistent homology that accounts for the topological features of WM lesions in a robust manner. Using computational geometry, \method{} takes the persistence of connected components into consideration, effectively filtering out the noisy WMH positives, resulting in a more accurate count of true lesions. We validated \method{} on the ISBI2015 longitudinal lesion segmentation dataset, where it produces significantly more accurate results than direct thresholding.


\end{abstract}
\section{Introduction}

White matter hyperintensities (WMH) are lesions that appear hyperintense on FLAIR MRI scans. WMH have many possible causes~\cite{barkhof2002imaging}, but the two most common are multiple sclerosis (MS)~\cite{lassmann2018multiple} and vascular disorders causing small vessel disease, often leading to stroke~\cite{vermeer2003silent}.  Furthermore, WMH have also been found to be associated with cognitive impairment and Alzheimer's disease (AD)~\cite{alber2019white}. Therefore, accurate quantification of WMH is valuable for the clinical assessment of these diseases and evaluation of potential treatment effect. 

Automated lesion segmentation is a key preprocessing step for reproducible, quantitative analysis of WMH -- particularly at large scale. Many image segmentation methods have been proposed for WMH. Representative classical methods include: BIANCA~\cite{griffanti2016bianca}, which relies on k-nearest neighbor classification; LST-LGA\cite{schmidt2012automated}, which uses a lesion growth algorithm; LST-LPA~\cite{schmidt2017bayesian}, which uses supervised logistic regression; atlas-based methods like Lesion-TOADS~\cite{shiee2010topology}; dictionary learning algorithms~\cite{weiss2013multiple}; or unsupervised Bayesian methods that are contrast-adaptive and rely on outlier detection~\cite{van2001automated}, such as 
BaMoS~\cite{sudre2015bayesian} or SAMSEG-lesion~\cite{cerri2021contrast}.

As in most medical image analysis domains, classical methods have been superseded by deep learning approaches~\cite{ma2022multiple}. Many of these methods rely on convolutional neural networks (CNNs) trained in a supervised fashion~\cite{brosch2016deep} -- possibly equipped with enhancements like positional encoding~\cite{ghafoorian2017location}, dedicated patch sampling strategies~\cite{guerrero2018white}, ensembles~\cite{manjon2018mri,li2018fully}, boundary losses to combat the large class imbalance~\cite{kervadec2019boundary}, or longitudinal strategies for jointly exploiting information from multiple timepoints and detect changes~\cite{vaidya2015longitudinal,kruger2020fully,gessert2020multiple}. Attempts have also been made to combine CNNs with ideas from the classical Bayesian literature to achieve resilience against changes in pulse sequence and image resolution~\cite{billot2021joint}.

Given segmentations, one can compute several quantitative metrics of interest, which are associated with patient outcomes. One such metric is the total lesion load (also known as lesion burden), which corresponds to the total amount of volume segmented as WMH -- typically computed from soft segmentations, i.e., by weighting the volume of each voxel by the lesion probability estimated by the automated algorithm at the given location. Lesion load has been shown to correlate with long-term outcomes in MS~\cite{popescu2013brain} and stroke~\cite{georgakis2019wmh}. Another important feature of WMH is their spatial distribution.
For example, the widespread Fazekas score for grading the amount of WHM in small vessel disease divides lesions into periventricular vs deep white matter~\cite{fazekas1987mr}. 

\begin{figure}[h]
\centering 
\subfigure[Original slice.]{
\includegraphics[width=0.23\textwidth]{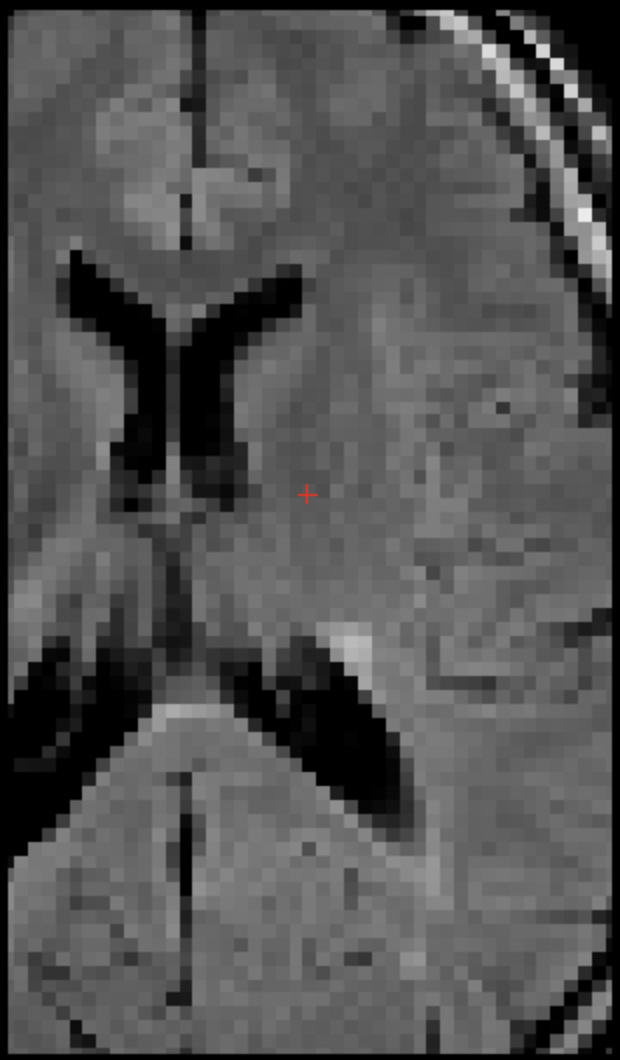}}
\hspace{-.08in}
\subfigure[Probability map.]{
\includegraphics[width=0.23\textwidth]{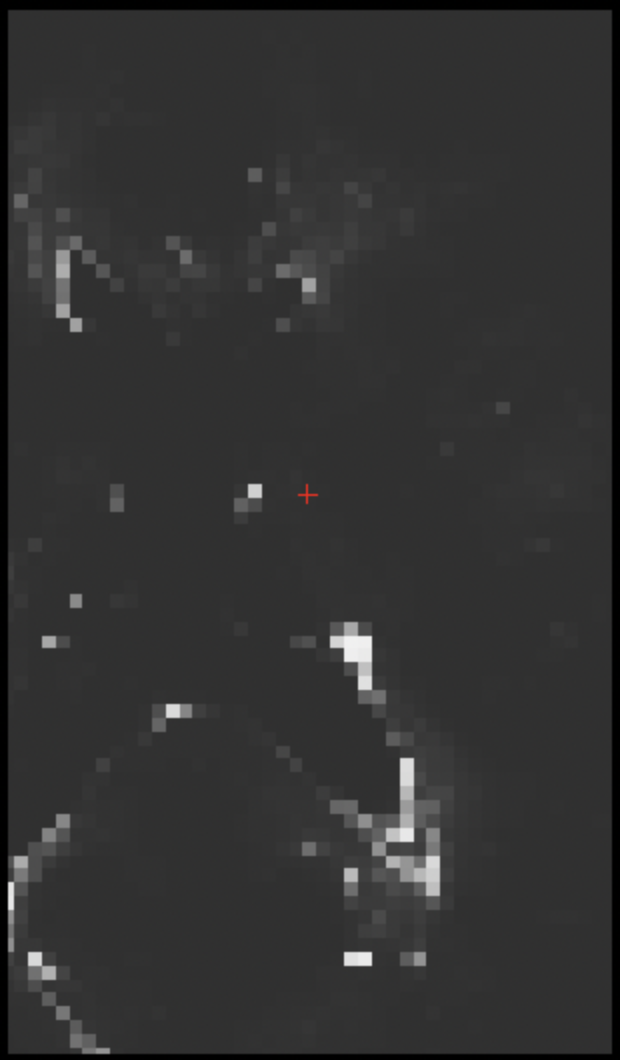}}
\hspace{-.08in}
\subfigure[Seg. under 0.3.]{
\includegraphics[width=0.23\textwidth]{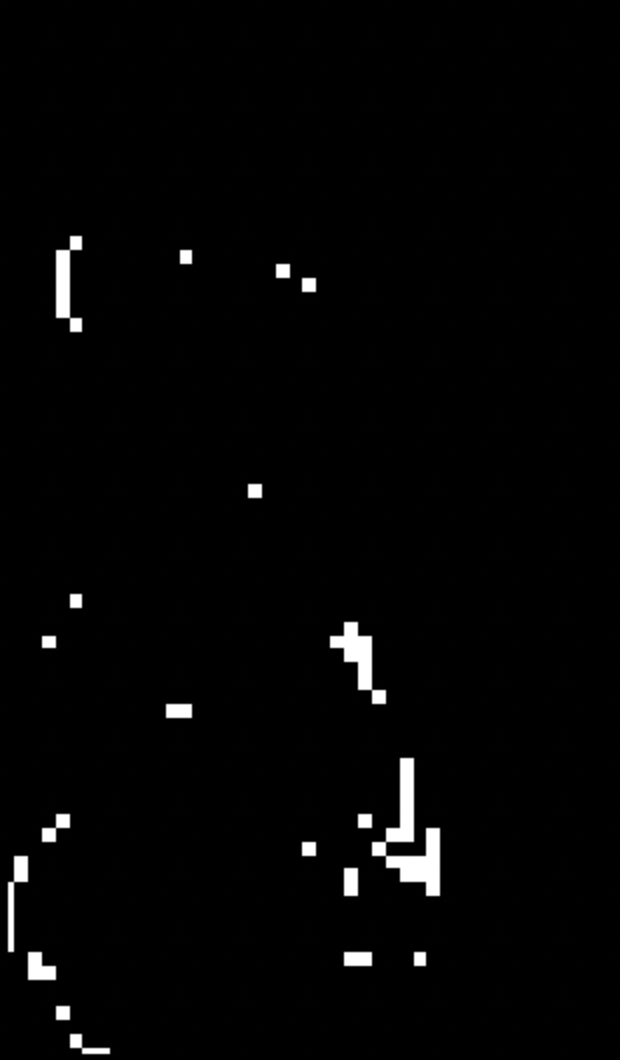}}
\hspace{-.08in}
\subfigure[Seg. under 0.7.]{
\includegraphics[width=0.23\textwidth]{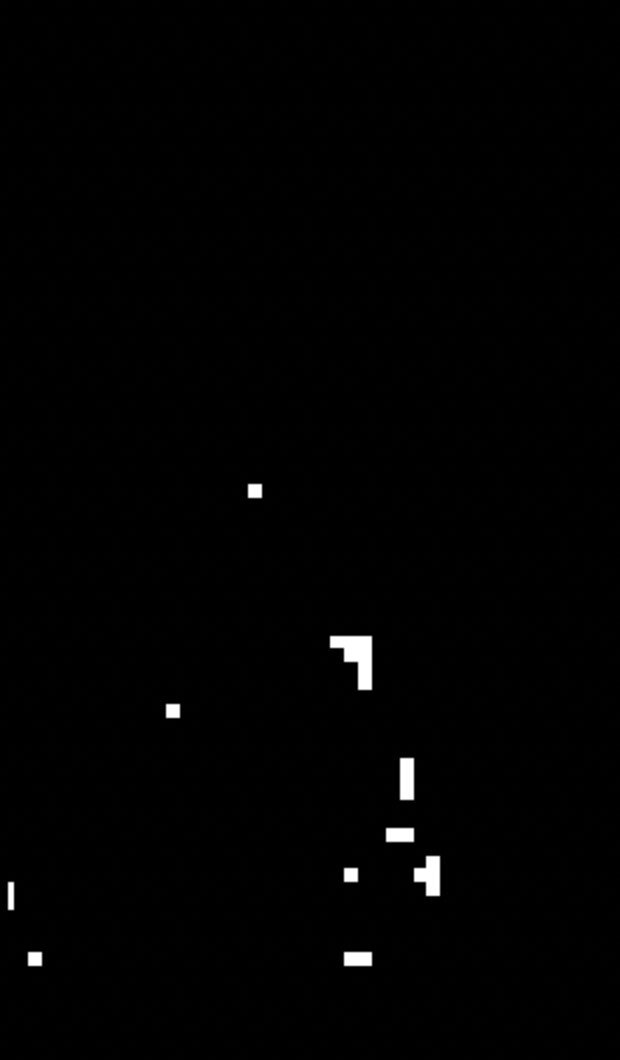}}
\caption{Motivation for \method{}: lesion counting from an MRI scan (a) based on direct thresholding of the soft probability map (b) is usually noisy and highly sensitive to the choice of the threshold (c,d).}
\label{fig:teaser}
\end{figure}

Another metric of interest that can be computed from WMH segmentations is the number of lesions. The appearance of new lesions (or enlargement of existing ones) is used to track the progression of MS in clinical practice, and has been shown to be predictive of disability~\cite{calabrese2012cortical,uher2017combining,treaba2019longitudinal}. However, counting lesions is an inherently difficult problem, with moderate inter-rater agreement~\cite{bozsik2022reproducibility,barkhof1999interobserver,zipoli2003interobserver}. While automated segmentation has the potential to curb this variability, counting lesions from a probabilistic segmentation in a robust fashion is not trivial. The standard approach consists of thresholding the lesion probability maps, computing connected components, and removing the smallest components (e.g., less than 3mm in the major axis~\cite{filippi2019assessment}) --  either by erosion or by volume thresholding (e.g., volumes under 10mm$^3$).
However, the lesion count is highly sensitive to the choice of threshold, as connected components are created (by splitting) and destroyed as the threshold increases (see Fig.~\ref{fig:teaser}). As the experiments below show, this lack of robustness leads to highly variable lesion counts -- which is particularly problematic in longitudinal data, as they obscure the real WMH progression.

In this article, we present \method{}, a novel method for robust lesion count using persistent homology (PH)~\cite{edelsbrunner2000topological,edelsbrunner2008persistent}. PH is a topological data analysis tool that has been applied to image segmentation of objects of known topology, both in medical (e.g., vessels, membranes, heart chambers~\cite{wu2017optimal,hu2019topology,hu2021topology,clough2020topological}) and natural images~\cite{chazal2013persistence}. Rather than leveraging PH in a supervised CNN as previous works, here we use it in an unsupervised fashion, to capture the full set of topological changes of the WMH probability map as a function of the threshold. This enables us to count lesions without having to explicitly threshold the probabilities, thus providing more robust estimates.


\section{Methods}

Our key innovation is to leverage the power of PH to count the number of lesions accurately. This is achieved through the insight that PH is robust to noisy information and able to capture the true signals effectively.

\subsection{Persistence-based Counting}
\label{sec:volume}
We first review the classical watershed algorithm for image segmentation, which is the basis of the proposed method. 

\begin{figure}[ht]
\centering 
\subfigure[1D watershed illustration.]{
\includegraphics[width=0.65\textwidth]{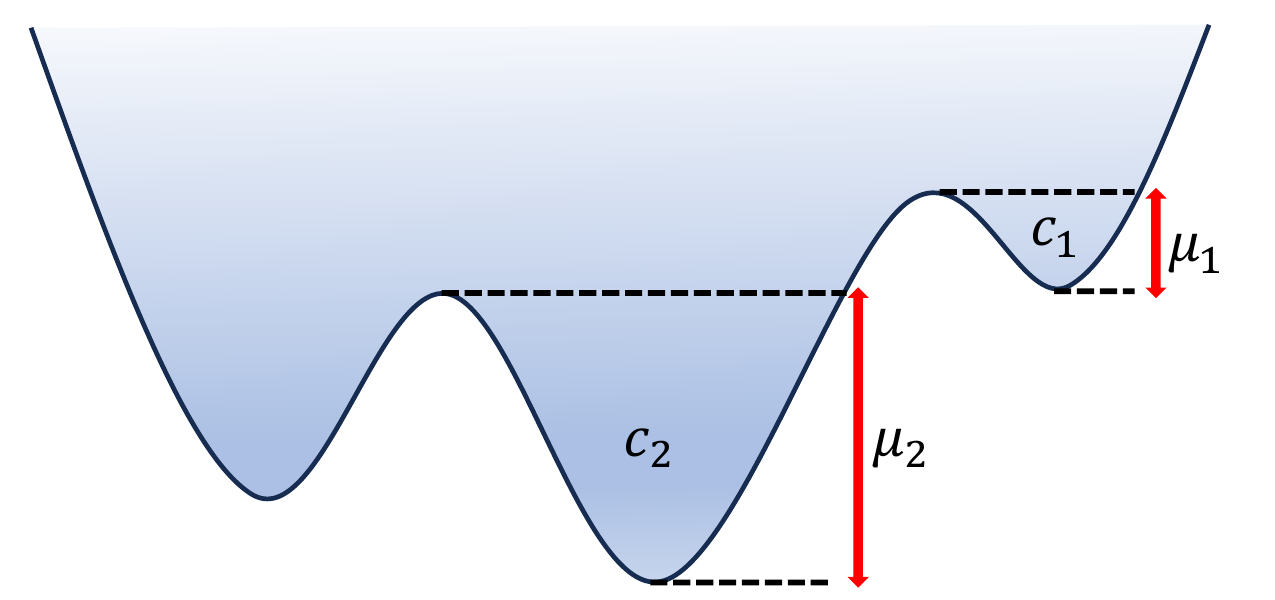}}
\hspace{-.08in}
\subfigure[Persistence diagram.]{
\includegraphics[width=0.33\textwidth]{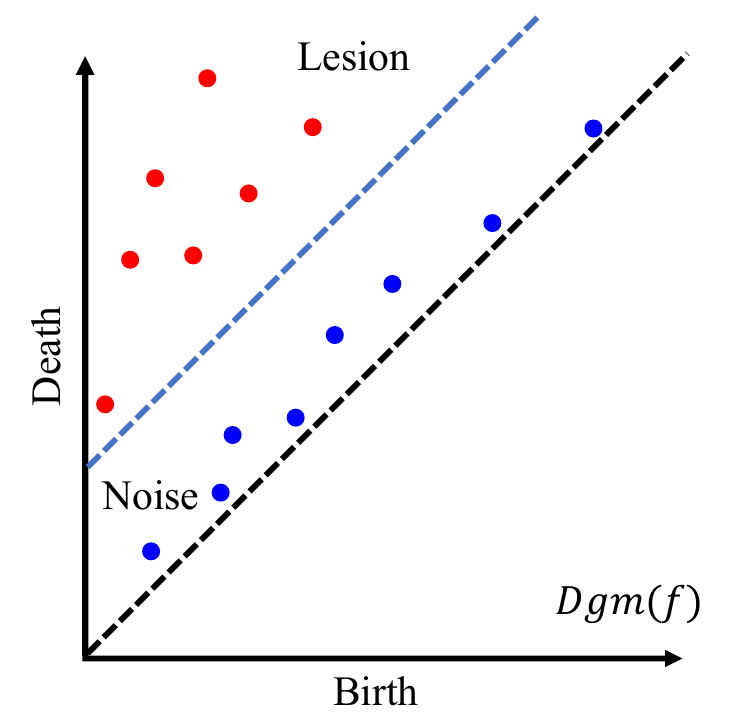}}
\caption{Illustration of the proposed \method{}. (\textbf{a})~Illustration of the watershed algorithm in 1D. As basins fill up, each is associated with a ``lifetime''. (\textbf{b})~Changes in connected components are captured by the persistence diagram as the threshold increases. }
\label{fig:topo}
\end{figure}

\myparagraph{Watershed algorithm.} By leveraging topographic information, the watershed algorithm divides a 2D/3D image into separated segments. It essentially treats the image as a terrain function (See Fig.~\ref{fig:topo}(a) for a 1D illustration), and identifies basins based on pixel/voxel intensity. Starting from local minima, the ``catchment'' basins fill up until region boundaries are reached. Each basin is then labeled as a separate region ($c_1$ and $c_2$ in Fig.~\ref{fig:topo}(a)), defining one single class / connected component. 

The result of the classical watershed algorithm relies heavily on the local minima. For example, $c_1$ and $c_2$ are both counted as valid connected components (i.e., lesions, in our context), regardless of the possibility that they may be noise. This poses significant challenges when quantifying lesions based on the segmentation algorithm, as every individual local minimum will be included in the count, irrespective of their size or persistence as the threshold varies.

To overcome this issue, we propose to leverage the tool of PH to capture the true lesion signals by suppressing the noisy ones. Specifically: instead of counting all the local minima, we seek to distinguish the `real' lesions from noise based on persistence under a progressively increasing threshold. Fig.~\ref{fig:topo}(a) shows an example of our intuition in 1D. Every basin/component is associated with a ``lifetime'' ($\mu_1$ for $c_1$, $\mu_2$ for $c_2$), which is a good indicator of how likely a connected component is to be a real lesion or not. Our method \method{} leverages PH  to address the issue. This is achieved by considering the persistence of connected components, effectively filtering out the noisy WMH positives, thus resulting in a more accurate count of true lesions. 

\myparagraph{Persistence-based counting.} For each soft probability map, we use a persistence diagram to capture the changes of the topological structure, as illustrated in Fig.~\ref{fig:topo}(b). Each dot in the persistence diagram corresponds to one single connected component, existing at a certain range of threshold values. To distinguish the `real' lesions from noise, we would like to decompose a diagram into `lesion' and `noise' parts. 

For a connected component, which corresponds to one dot in the persistence diagram, its lifetime is the persistence of the corresponding dot, i.e., the difference between its death and birth time: $per(p) = death(p) - birth(p)$. And the `persistence' of a dot in Fig.~\ref{fig:topo}(b) is truly the lifetime ($\mu_1$ and $\mu_2$) in Fig.~\ref{fig:topo}(a).
Persistence is a good metric indicating how likely a connected component is to be a real lesion or not: the greater the persistence, the longer the connected component exists through the whole ``filtration'', and the more likely the connected component is a real `lesion'. In contrast, the connected components with low persistence are more likely to be `noise'.

As a result, we can filter out the real lesions based on the persistence using a predetermined threshold $\theta$. Driven by this, we propose a novel algorithm called \method{} for white matter hyperintensities in brain MRI. The details are illustrated in Alg.~\ref{alg:ph_watershed}.
As we show in the experiments below, the proposed persistence-based counting algorithm is more robust and results in more accurate counting, especially for noisy inputs. The results also show that the choice of threshold: \textit{(i)}~has a much lesser impact on the variability of the lesion count than a threshold taken directly on the probability maps; and \textit{(ii)}~has very little impact on the longitudinal counts.

\begin{table}[ht]
\begin{algorithm}[H]
\caption{\method{}}
\label{alg:ph_watershed}
\KwIn{A 3D soft lesion probability map, and a threshold $\theta$}
\KwOut{Number of lesions}
\textbf{Definition}: $G =(V, E)$ denote a graph; 
$f(v)$ is the intensity value of node $v$; lower\_star$(v)$ = $\{(u,v) \in E| f(u) < f(v) \}$; 
$cc(v)$ is the connected component id of node $v$.  
\begin{algorithmic}[1] 
\STATE PD =$\emptyset$; Build the proximity graph (6-connectivity) for 3D image; 

\STATE $U = V$ sorted according to $f(v)$; $T$ a sub-graph, which includes all the nodes and edges whose value $<t$.
\FOR {$v$ in $U$}
\STATE $t= f(v)$, $T = T + \{v\}$

\FOR {$(u,v)$ in lower\_star($v$)}
\item Assert $u \in T$

\IF{$cc(u) = cc(v)$}
\item Continue
\ELSE
\item younger\_cc = $\argmaxA_{w=cc(u), cc(v)} f(w)$
\item older\_cc = $\argminA_{w=cc(u), cc(v)} f(w)$

\item \textit{pers = $t$-$f$(younger\_cc)}
\IF{pers <= $\theta$}
\FOR {$w$ in younger\_cc} 
\item $cc(w)$ = older\_cc
\ENDFOR
\ENDIF

\item PD = PD + ($f$(younger\_cc), $t$)
\ENDIF
\ENDFOR
\ENDFOR

\STATE \textbf{return} $\#$ of lesions = len(cc).
\end{algorithmic}
\end{algorithm}
\end{table}

\subsection{Optimal threshold selection}
\label{optimal_thresh}
Let us assume the availability of $N$ training samples, and that the $i$-th sample has $T_{i}$ time points. For a set of thresholds $\{\theta_{j}\}$, sample $i$ has $y_{ij1}$, $y_{ij2}$, ..., $y_{ij{T_{i}}}$ number of lesions for time point $1$, $2$, ..., $T_{i}$, respectively. The optimal value of threshold depends on the dataset. To find this optimal value, we propose a  supervised and an unsupervised approach, depending on whether ground truth labels are available for some scans or not. 

\myparagraph{Supervised approach.} If ground truth labels are given, we use a supervised approach to select the optimal $\theta$. Let's use $\tilde{y}_{i1}$, $\tilde{y}_{i2}$, ..., $\tilde{y}_{i{T_{i}}}$ to denote the ground truth number of lesions for sample $i$ at time point $1$, $2$, ..., $T_{i}$, respectively. Then, we simply pick the threshold that minimizes the sum of absolute errors over all time points of all  training samples:
\begin{equation}
\theta^{*} = {arg\,min}_\theta \sum\limits_{i=1, ... , N; t = 1, ..., T_i } (\tilde{y}_{it} - y_{ijt})^2.
    \label{eq:supervised}
\end{equation}

\myparagraph{Unsupervised approach.} If ground truth labels are not available, we utilize an unsupervised method to find the optimal $\theta$. Specifically, we fit a linear model to the number of lesions over time. For sample $i$ under a specific threshold $\theta_j$, we have:
\begin{equation}
    \hat{y}_{ijt} = a * t + b + \epsilon_{ijt}, \quad t = 1, 2, ..., T_{i}.
\end{equation}
where $\hat{y}_{ijt}$ is the regressed number of lesions at threshold $j$, time point $t$ for sample $i$ and $\epsilon_{ijt}$ models the errors. We use least squares (L2 norm) to fit the linear curve. Different from the supervised setting, we minimize the following term:
\begin{equation}
\theta^{*} = arg\,min \sum\limits_{i=1, ... , N; t = 1, ..., T_i } (\hat{y}_{ijt} - y_{ijt})^2.
\label{eq:unsupervised}
\end{equation}

The optimal $\theta$ can be found through Eq.~(\ref{eq:supervised}) or Eq.~(\ref{eq:unsupervised}) under supervised or unsupervised settings, respectively. The optimal $\theta$ is then applied to the test set.


\section{Experiments and Results}
\myparagraph{Datasets.} To validate the effectiveness of the proposed method, we use the training subset of a longitudinal, multi-modality dataset of WMH (ISBI15~\cite{carass2017longitudinal}), for which manual longitudinal segmentations are available. Each subject has 4 or 5 timepoints. We used the FLAIR scans, resampled to 1mm isotropic resolution -- which is the native resolution of the manual segmentations. 

\myparagraph{Automated segmentation.} We used SAMSEG-lesion~\cite{cerri2021contrast} as the automated segmentation method to obtain the soft probabilities from the original scans. We used SAMSEG-lesion because it's adaptive to contrast and therefore generalizes very well to ISBI2015. The output was a soft segmentation of WMH at 1mm isotropic resolution.

\myparagraph{Implementation details.} Our algorithm is computationally expensive, due to the need to monitor connected components at small threshold increases. To reduce its computational burden, we aggressively crop the imaging volumes around the brain, and downsample them to  2mm $\times$ 2mm $\times$ 2mm resolution. This yields volumes of approximately  $40 \times 80 \times 40$ voxels, which allows our Python implementation to process a volume in approximately 50sec on a modern desktop.

\myparagraph{Baseline.} We use direct thresholding (which is the current standard in clinical practice) as the baseline to show the effectiveness of the proposed method. Specifically, we do direct thresholding on the obtained soft probability map, and then count the number of connected components to obtain the number of lesions. We note that we do not remove small components as the voxel size is 8mm$^3$ which is approximately equal to the threshold suggested in~\cite{georgakis2019wmh}.

\myparagraph{Evaluation.} For the baseline method, we gradually increase the threshold from 0.1 to 1 (step size is 0.1), and plot the resulting lesion count vs time curves. For the proposed \method{} method, we similarly increase the persistence threshold ($\theta$ in Alg.~\ref{alg:ph_watershed}) from 0 to 0.04 (step size is 0.004) to plot the same curves. To evaluate the algorithm quantitatively, we studied the absolute error in lesion count with respect to the ground truth, computed with five-fold cross-validation.

\begin{figure}[t]
\centering
\subfigure[Subject 2 (direct thresholding).]{
\includegraphics[width=0.49\textwidth]{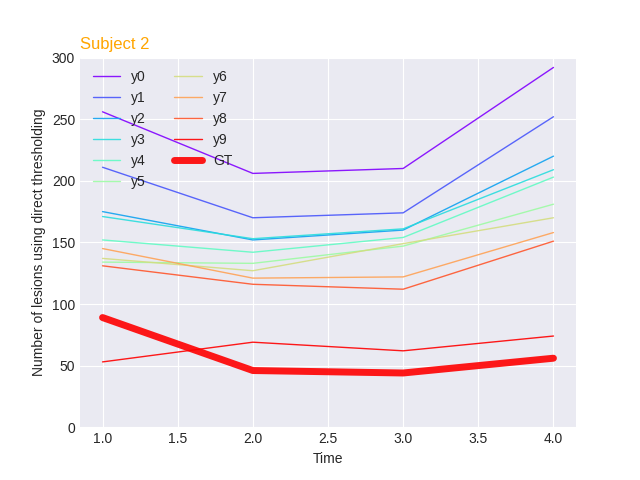}}
\hspace{-.08in}
\subfigure[Subject 2 (\method{}).]{
\includegraphics[width=0.49\textwidth]{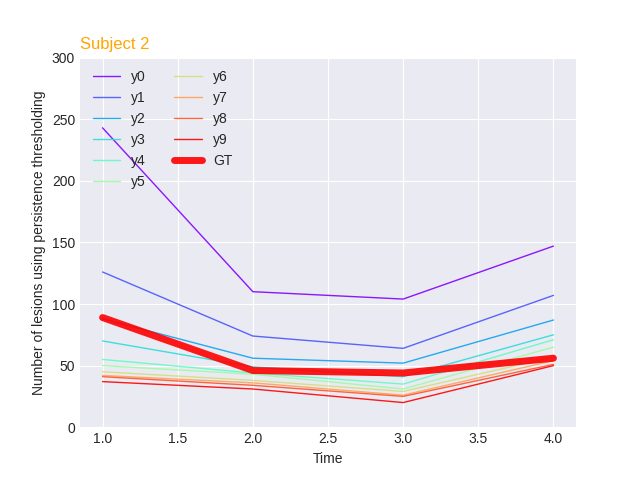}}

\subfigure[Subject 4 (direct thresholding).]{
\includegraphics[width=0.49\textwidth]{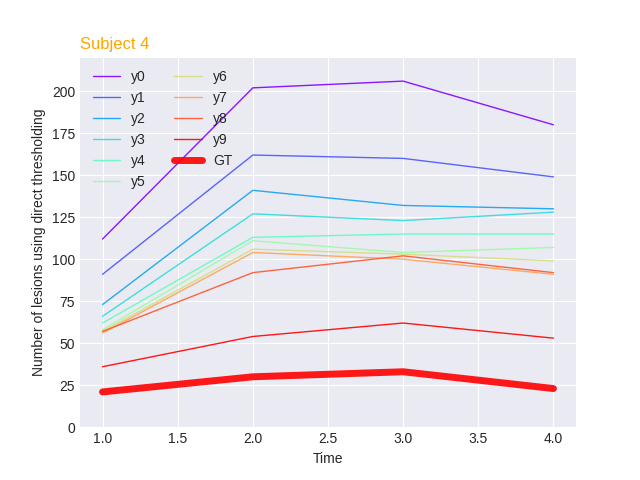}}
\hspace{-.08in}
\subfigure[Subject 4 (\method{}).]{
\includegraphics[width=0.49\textwidth]{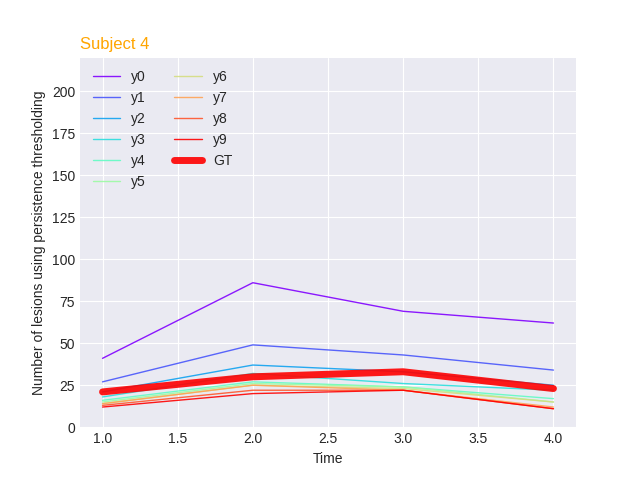}}
\caption{Lesion count vs time for Subjects 2 and 4 of  ISBI2015 using different thresholds (purple=more liberal; red=more conservative), for direct thresholding and \method{}. The thick red line corresponds to the ground truth count derived from manual segmentations.}
\label{Spaghetti}
\end{figure}

\begin{figure}[t]
\centering 
\subfigure[Direct thresholding.]{
\includegraphics[width=0.31\textwidth]{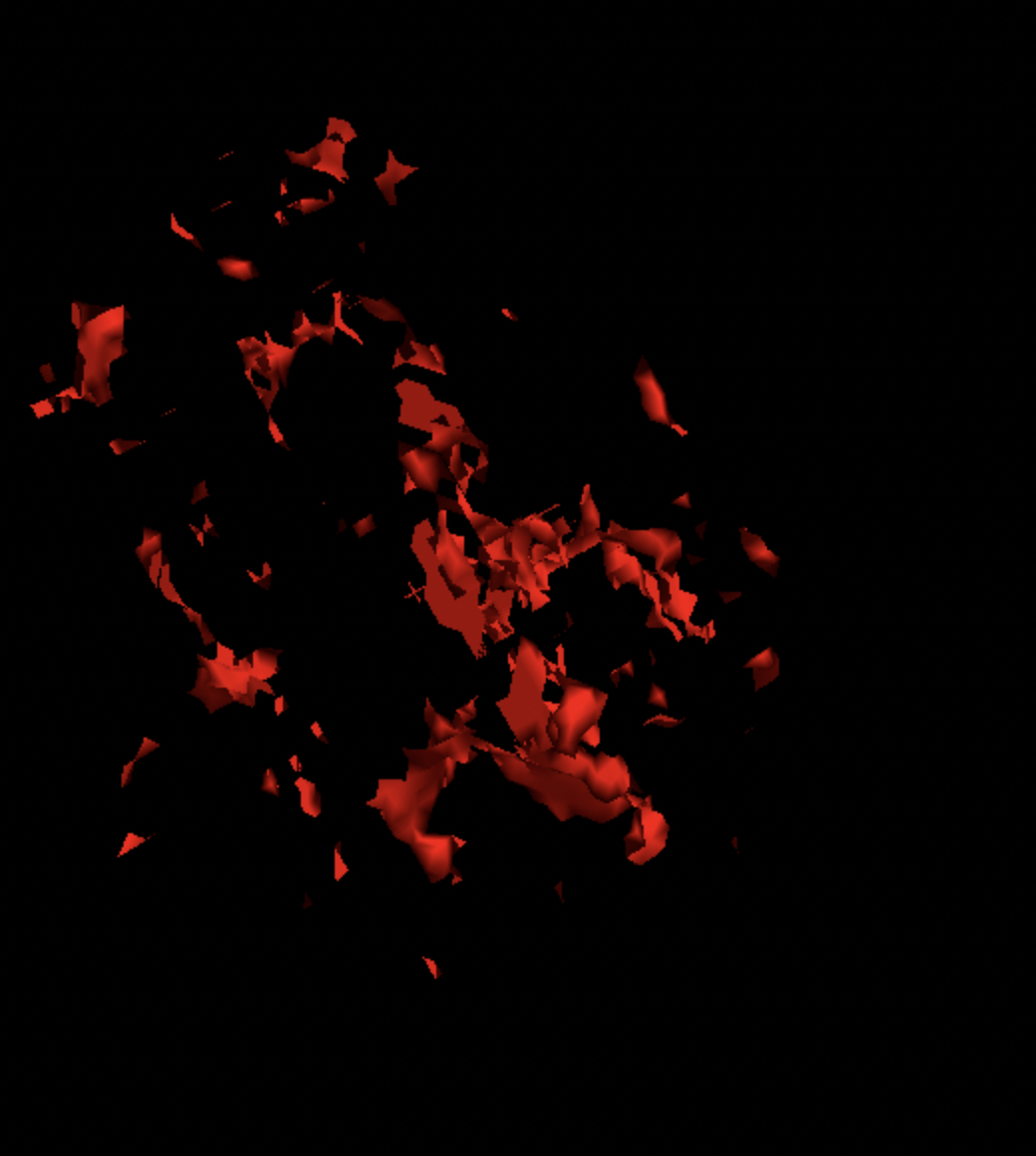}}
\hspace{-.08in}
\subfigure[\method{}.]{
\includegraphics[width=0.31\textwidth]{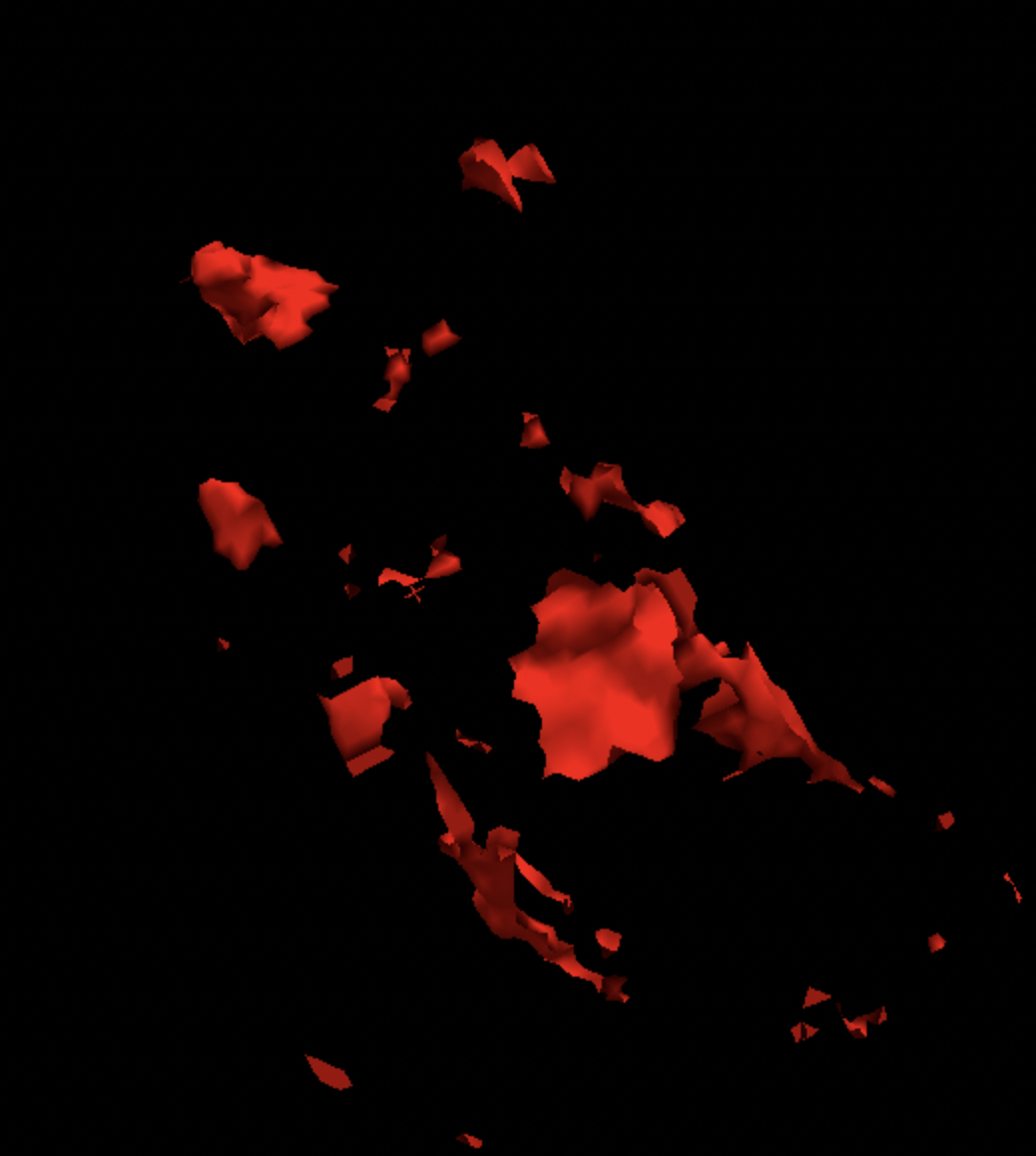}}
\hspace{-.08in}
\subfigure[GT.]{
\includegraphics[width=0.31\textwidth]{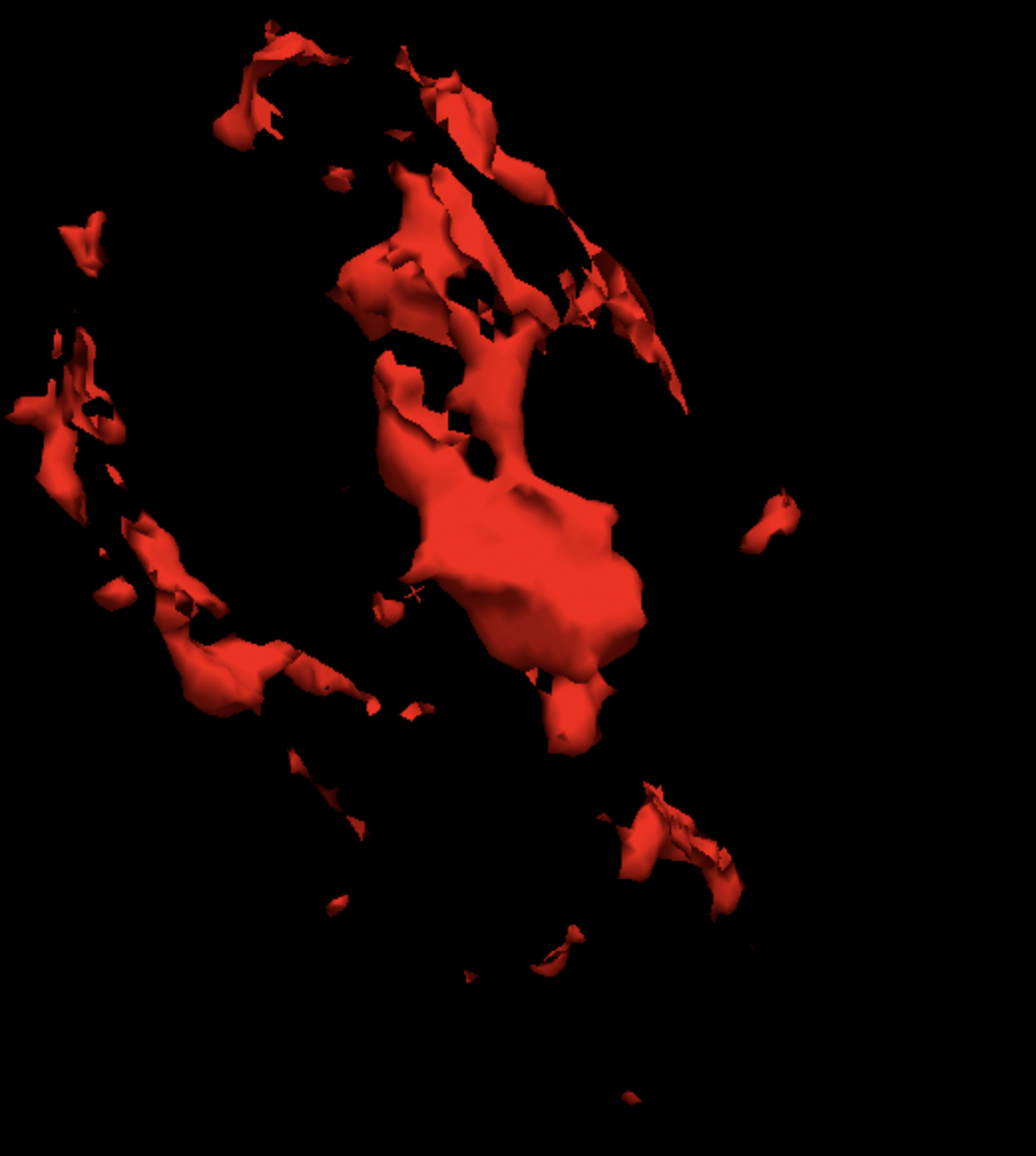}}
\hspace{-.08in}
\caption{3D rendering of WMH for a sample subject. (a)~Direct thresholding. (b)~\method{}. (c)~Ground truth. We note that a segmentation like (b) cannot be obtained by thresholding the probability map at any level, as it is based on persistence.}
\label{fig:seg}
\end{figure}

\myparagraph{Results.} 
Fig.~\ref{Spaghetti} shows the temporal evolution of the number of lesions for two sample subjects in the dataset (Subjects 2 and 4), while Tab.~\ref{tab:method} shows the average errors in lesion counts, for the supervised and unsupervised choices of threshold. 
Qualitatively speaking, Fig.~\ref{Spaghetti} illustrates two major advantages of \method{} with respect to the baseline. First, and most obvious: the error in the lesion count is much lower. This is also apparent from Tab.~\ref{tab:method}, with strongly significant reductions in the error rate (despite the small sample size) -- particularly for the unsupervised approach. The second improvement is the much lower dependence on the threshold. Direct thresholding of probability maps is very noisy, as illustrated in Fig.~\ref{fig:seg}(a), which leads to a wide range of trajectories in Fig.~\ref{Spaghetti}(a) and Fig.~\ref{Spaghetti}(c). This variability is also noticeable in Tab.~\ref{tab:method}, where the thresholds obtained by the supervised and unsupervised approaches yield very different error rates (one almost as twice as the other). Our method, on the other one, thresholds the \emph{persistence}, yielding curves that are much closer to each other (Figs.~\ref{Spaghetti}(b) and ~\ref{Spaghetti}(d)), segmentations closer to the ground truth (Fig.~\ref{fig:seg}(b)), and errors that are less dependent on the chosen method to determine the threshold (Tab.~\ref{tab:method}).

\setlength{\tabcolsep}{6pt}
\begin{table}[ht]
\begin{center}
\caption{Mean absolute errors for lesion count (in number of lesions). The p-values are for two-tailed t-tests comparing the error rates of \method{} and direct thresholding.}
\label{tab:method}
\begin{tabular}{c|cc}
\hline
Method & Mean absolute errors  & $p$-value \\
\hline
Direct thresholding - Unsupervised & 43.19  & \\ 
\method{} - Unsupervised &  \textbf{16.81} & $3.48 * 10^{-5}$ \\
\hline
Direct thresholding - Supervised & 26.86 &  \\
\method{} - Supervised &   \textbf{19.14} & $1.07 * 10^{-3}$ \\
\hline
\end{tabular}
\end{center}
\end{table}


\section{Conclusion}
We have presented \method{}, a PH method for counting WMH  that accounts for their topological features in a robust manner. \method{} yields much lower errors than the standard thresholding currently used in clinical practice. \method{} also has limitations, notably its high computational complexity; future work will seek to address this issue. We believe that \method{} has great potential in increasing the accuracy of WMH quantification for the clinical assessment of several diseases and for the evaluation of the effect of treatments.

\section*{Acknowledgement}
Supported by NIH grants 1RF1MH123195, 1R01AG070988, 1RF1AG080371, 1UM1MH130981. OP was supported by a grant from Lundbeckfonden (grant number R360–2021–395).

\bibliographystyle{splncs04}
\bibliography{ref}

\begin{thebibliography}{10}
\providecommand{\url}[1]{\texttt{#1}}
\providecommand{\urlprefix}{URL }
\providecommand{\doi}[1]{https://doi.org/#1}

\bibitem{alber2019white}
Alber, J., Alladi, S., Bae, H.J., et~al.: White matter hyperintensities in vascular contributions to cognitive impairment and dementia ({VCID}): Knowledge gaps and opportunities. Alzheimer's \& Dementia: Translational Research \& Clinical Interventions  (2019)

\bibitem{barkhof1999interobserver}
Barkhof, F., Filippi, M., Waesberghe, V., et~al.: Interobserver agreement for diagnostic {MRI} criteria in suspected multiple sclerosis. Neuroradiology  (1999)

\bibitem{barkhof2002imaging}
Barkhof, F., Scheltens, P.: Imaging of white matter lesions. Cerebrovascular Diseases  (2002)

\bibitem{billot2021joint}
Billot, B., Cerri, S., Leemput, V., et~al.: Joint segmentation of multiple sclerosis lesions and brain anatomy in {MRI} scans of any contrast and resolution with {CNNs}. In: ISBI (2021)

\bibitem{bozsik2022reproducibility}
Bozsik, B., T{\'o}th, E., Poly{\'a}k, I., et~al.: Reproducibility of lesion count in various subregions on {MRI} scans in multiple sclerosis. Frontiers in Neurology  (2022)

\bibitem{brosch2016deep}
Brosch, T., Tang, L.Y., Yoo, Y., et~al.: Deep {3D} convolutional encoder networks with shortcuts for multiscale feature integration applied to multiple sclerosis lesion segmentation. TMI  (2016)

\bibitem{calabrese2012cortical}
Calabrese, M., Poretto, V., Favaretto, A., et~al.: Cortical lesion load associates with progression of disability in multiple sclerosis. Brain  (2012)

\bibitem{carass2017longitudinal}
Carass, A., Roy, S., Jog, A., et~al.: Longitudinal multiple sclerosis lesion segmentation: resource and challenge. NeuroImage  (2017)

\bibitem{cerri2021contrast}
Cerri, S., Puonti, O., Meier, D.S., et~al.: A contrast-adaptive method for simultaneous whole-brain and lesion segmentation in multiple sclerosis. Neuroimage  (2021)

\bibitem{chazal2013persistence}
Chazal, F., Guibas, L.J., Oudot, S.Y., et~al.: Persistence-based clustering in riemannian manifolds. Journal of the ACM  (2013)

\bibitem{clough2020topological}
Clough, J.R., Byrne, N., Oksuz, I., et~al.: A topological loss function for deep-learning based image segmentation using persistent homology. TPAMI  (2020)

\bibitem{edelsbrunner2008persistent}
Edelsbrunner, H., Harer, J., et~al.: Persistent homology -- a survey. Contemporary mathematics  (2008)

\bibitem{edelsbrunner2000topological}
Edelsbrunner, H., Letscher, D., Zomorodian, A.: Topological persistence and simplification. In: FOCS (2000)

\bibitem{fazekas1987mr}
Fazekas, F., Chawluk, J.B., Alavi, A., et~al.: {MR} signal abnormalities at {1.5 T} in {Alzheimer's} dementia and normal aging. American Journal of Neuroradiology  (1987)

\bibitem{filippi2019assessment}
Filippi, M., Preziosa, P., Banwell, B.L., et~al.: Assessment of lesions on magnetic resonance imaging in multiple sclerosis: practical guidelines. Brain  (2019)

\bibitem{georgakis2019wmh}
Georgakis, M.K., Duering, M., Wardlaw, J.M., et~al.: W{MH} and long-term outcomes in ischemic stroke: a systematic review and meta-analysis. Neurology  (2019)

\bibitem{gessert2020multiple}
Gessert, N., Kr{\"u}ger, J., Opfer, R., et~al.: Multiple sclerosis lesion activity segmentation with attention-guided two-path cnns. Computerized Medical Imaging and Graphics  \textbf{84},  101772 (2020)

\bibitem{ghafoorian2017location}
Ghafoorian, M., Karssemeijer, N., Heskes, T., et~al.: Location sensitive deep convolutional neural networks for segmentation of white matter hyperintensities. Scientific Reports  (2017)

\bibitem{griffanti2016bianca}
Griffanti, L., Zamboni, G., Khan, A., et~al.: {BIANCA} ({Brain} {Intensity} {AbNormality} {Classification} {Algorithm}): A new tool for automated segmentation of white matter hyperintensities. Neuroimage  (2016)

\bibitem{guerrero2018white}
Guerrero, R., Qin, C., Oktay, O., et~al.: White matter hyperintensity and stroke lesion segmentation and differentiation using convolutional neural networks. NeuroImage: Clinical  (2018)

\bibitem{hu2019topology}
Hu, X., Li, F., Samaras, D., et~al.: Topology-preserving deep image segmentation. NeurIPS  (2019)

\bibitem{hu2021topology}
Hu, X., Wang, Y., Fuxin, L., et~al.: Topology-aware segmentation using discrete morse theory. ICLR  (2021)

\bibitem{kervadec2019boundary}
Kervadec, H., Bouchtiba, J., Desrosiers, C., et~al.: Boundary loss for highly unbalanced segmentation. In: MIDL (2019)

\bibitem{kruger2020fully}
Kr{\"u}ger, J., Opfer, R., Gessert, N., et~al.: Fully automated longitudinal segmentation of new or enlarged multiple sclerosis lesions using {3D} convolutional neural networks. NeuroImage: Clinical  (2020)

\bibitem{lassmann2018multiple}
Lassmann, H.: Multiple sclerosis pathology. Cold Spring Harbor perspectives in medicine  (2018)

\bibitem{li2018fully}
Li, H., Jiang, G., Zhang, J., et~al.: Fully convolutional network ensembles for white matter hyperintensities segmentation in {MR} images. NeuroImage  (2018)

\bibitem{ma2022multiple}
Ma, Y., Zhang, C., Cabezas, M., et~al.: Multiple sclerosis lesion analysis in brain magnetic resonance images: techniques and clinical applications. JBHI  (2022)

\bibitem{manjon2018mri}
Manj{\'o}n, J.V., Coup{\'e}, P., Raniga, P., et~al.: {MRI} white matter lesion segmentation using an ensemble of neural networks and overcomplete patch-based voting. Computerized Medical Imaging and Graphics  (2018)

\bibitem{popescu2013brain}
Popescu, V., Agosta, F., Hulst, H.E., et~al.: Brain atrophy and lesion load predict long term disability in multiple sclerosis. Journal of Neurology, Neurosurgery \& Psychiatry  (2013)

\bibitem{schmidt2017bayesian}
Schmidt, P.: Bayesian inference for structured additive regression models for large-scale problems with applications to medical imaging. Ph.D. thesis, lmu (2017)

\bibitem{schmidt2012automated}
Schmidt, P., Gaser, C., Arsic, M., et~al.: An automated tool for detection of flair-hyperintense white-matter lesions in multiple sclerosis. Neuroimage  (2012)

\bibitem{shiee2010topology}
Shiee, N., Bazin, P.L., Ozturk, A., et~al.: A topology-preserving approach to the segmentation of brain images with multiple sclerosis lesions. NeuroImage  (2010)

\bibitem{sudre2015bayesian}
Sudre, C.H., Cardoso, M.J., Bouvy, W.H., et~al.: Bayesian model selection for pathological neuroimaging data applied to white matter lesion segmentation. TMI  (2015)

\bibitem{treaba2019longitudinal}
Treaba, C.A., Granberg, T.E., Sormani, M.P., et~al.: Longitudinal characterization of cortical lesion development and evolution in multiple sclerosis with 7.0-{T} {MRI}. Radiology  (2019)

\bibitem{uher2017combining}
Uher, T., Vaneckova, M., Sobisek, L., et~al.: Combining clinical and magnetic resonance imaging markers enhances prediction of 12-year disability in multiple sclerosis. Multiple Sclerosis Journal  (2017)

\bibitem{vaidya2015longitudinal}
Vaidya, S., Chunduru, A., Muthuganapathy, R., et~al.: Longitudinal multiple sclerosis lesion segmentation using {3D} convolutional neural networks. Proceedings of the 2015 longitudinal multiple sclerosis lesion segmentation challenge  (2015)

\bibitem{van2001automated}
Van~Leemput, K., Maes, F., Vandermeulen, D., et~al.: Automated segmentation of multiple sclerosis lesions by model outlier detection. TMI  (2001)

\bibitem{vermeer2003silent}
Vermeer, S.E., Hollander, M., van Dijk, E.J., et~al.: Silent brain infarcts and white matter lesions increase stroke risk in the general population: the rotterdam scan study. Stroke  (2003)

\bibitem{weiss2013multiple}
Weiss, N., Rueckert, D., Rao, A.: Multiple sclerosis lesion segmentation using dictionary learning and sparse coding. In: MICCAI (2013)

\bibitem{wu2017optimal}
Wu, P., Chen, C., Wang, Y., et~al.: Optimal topological cycles and their application in cardiac trabeculae restoration. In: MICCAI (2017)

\bibitem{zipoli2003interobserver}
Zipoli, V., Portaccio, E., Siracusa, G., et~al.: Interobserver agreement on {Poser’s} and the new {McDonald’s} diagnostic criteria for multiple sclerosis. Multiple Sclerosis Journal  (2003)

\end{thebibliography}

\end{document}